# Anisotropic Optical Conductivity Accompanied by a Small Energy Gap in One-Dimensional Thermoelectric Telluride Ta$_4$SiTe$_4$

Fumiya Matsunaga[1], Yoshihiko Okamoto[1,2,*], Yasunori Yokoyama[1], Kanji Takehana[3], Yasutaka Imanaka[3], Yuto Nakamura[1], Hideo Kishida[1], Shoya Kawano[4], Kazuyuki Matsuhira[4], and Koshi Takenaka[1]

[1]Department of Applied Physics, Nagoya University, Nagoya 464-8603, Japan
[2]Institute for Solid State Physics, University of Tokyo, Kashiwa 277-8581, Japan
[3] National Institute for Materials Science (NIMS), Sakura 3-13, Tsukuba 305-0003, Japan.
[4]Graduate School of Engineering, Kyushu Institute of Technology, Kitakyushu 804-8550, Japan.

We investigated the optical properties of single crystals of one-dimensional telluride Ta$_4$SiTe$_4$, which shows high thermoelectric performance below room temperature. Optical conductivity estimated from reflectivity spectra indicates the presence of a small energy gap of 0.1–0.15 eV at the Fermi energy. At the lowest energy, optical conductivity along the Ta$_4$SiTe$_4$ chain is an order of magnitude higher than that perpendicular to this direction, reflecting the anisotropic electron conduction in Ta$_4$SiTe$_4$. These results indicate that coexistence of a very small band gap and anisotropic electron conduction is a promising strategy to develop a high-performance thermoelectric material for low temperature applications.

There are great expectations for thermoelectric energy conversion between thermal and electrical energy, which can be used for energy harvesting and local cooling. At present, thermoelectric energy conversion has been put to practical use in Peltier cooling at around room temperature using Bi$_2$Te$_3$-based materials, and in radioisotope thermoelectric generators using PbTe or SiGe-based materials. A new material that exhibits much higher performance at room temperature will open an avenue for practical use of energy harvesting, which obtains electrical energy from the temperature differences around us. In addition, a new material that exhibits high performance below −100 °C, where Bi$_2$Te$_3$-based materials cannot be used, will lead to Peltier cooling and precision temperature control at low temperatures. In recent years, the development of new materials for high-temperature applications has been remarkable. PbTe with hierarchical architectures, AgPb$_m$SbTe$_{2+m}$, and SnSe have been reported to exhibit very low thermal conductivity $\kappa$, resulting in a large dimensionless figure of merit $ZT = S^2T/\rho\kappa$ exceeding 2 at high temperatures, where $S$, $\rho$, and $T$ are Seebeck coefficient, electrical resistivity, and temperature, respectively [1–4]. In contrast, there were far fewer candidate materials, such as Bi$_{1-x}$Sb$_x$ and CsBi$_4$Te$_6$, for low-temperature applications [5–7]. This reflects the fact that at low temperatures, a reduction in $\kappa$ alone is not sufficient to achieve high thermoelectric performance, but an increase in $S$ and a reduction in $\rho$ are also essential.

Recently, one-dimensional telluride Ta$_4$SiTe$_4$ and its Mo-

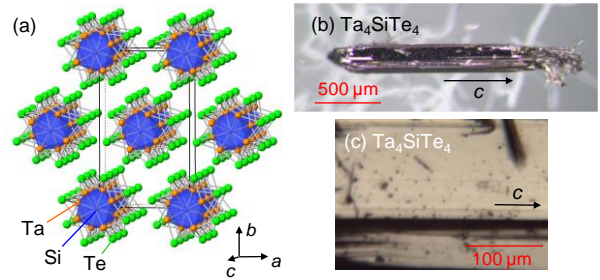

Figure 1. (a) Crystal structure of Ta$_4$SiTe$_4$. The orthorhombic unit cell is indicated by solid lines. (b) A single crystal of Ta$_4$SiTe$_4$. (c) A typical crystal surface used in optical reflectivity measurements.

doped samples were reported to exhibit a very large |$S$| with sufficiently small $\rho$ for thermoelectric materials over wide temperatures from 50 K to room temperature [8]. Ta$_4$SiTe$_4$ has a one-dimensional crystal structure with the orthorhombic *Pbam* space group, consisting of Ta$_4$SiTe$_4$ chains loosely bounded by van der Waals interactions between Te atoms, as shown in Fig. 1(a) [9,10]. In Ta$_4$SiTe$_4$, Ta$_4$SiTe$_4$ chains lie parallel to the *c*-axis, forming an almost perfect triangular lattice in the *ab* plane, and making the material almost isotropic in this plane. Whisker crystals with a length of several millimeters and a thickness of at most 10 μm were synthesized, and $\rho$ and $S$ along the whisker, //*c*, were measured [8]. Furthermore, 0.1%–0.2% Mo-doped whiskers exhibited a large negative Seebeck coefficient of |$S$| ~ 300 μV K$^{-1}$ and a small $\rho$ = 1 mΩ cm at 220–280 K, resulting in a

huge power factor $P = S^2/\rho$ of 170 μW cm$^{-1}$K$^{-2}$. This $P$ is more than four times the room-temperature value of Bi$_2$Te$_3$-based practical materials.

Since the above report, research on this system as a thermoelectric material has progressed. The whisker crystals of chemically doped Nb$_4$SiTe$_4$, which is a 4$d$ analogue of Ta$_4$SiTe$_4$, and the solid solution between Ta$_4$SiTe$_4$ and Nb$_4$SiTe$_4$ also showed a large $P$ exceeding those in practical materials [11,12]. Moreover, $p$-type whisker crystals were obtained by Ti doping to the Ta site [13]. The power factor of the $p$-type whiskers reached a maximum value of 60 μW cm$^{-1}$K$^{-2}$ and exceeded the practical level between 100 K and room temperature. Furthermore, the thermoelectric properties of sintered Ta$_4$SiTe$_4$ samples prepared by a cold press method and of a flexible composite of Ta$_4$SiTe$_4$ whiskers and an organic conductor were investigated [14,15].

In contrast, the physical background behind the realization of huge $P$ below room temperature in this system has not been clarified experimentally. First-principles calculations showed that Ta$_4$SiTe$_4$ and Nb$_4$SiTe$_4$ have a one-dimensional band structure, in which a small band gap opens at the Dirac point due to strong spin–orbit coupling [8,10,16]. It is natural to believe that this characteristic band structure plays an important role in achieving huge $P$ at low temperatures. However, there have been few experimental studies on the electronic state and the correlation between the electronic state and thermoelectric properties of this system. The whisker morphology of the synthesized samples has hampered experimental studies. In this Letter, we report reflectivity spectra of Ta$_4$SiTe$_4$ single crystals measured over a wide energy range using linearly polarized light oscillating parallel or perpendicular to the Ta$_4$SiTe$_4$ chains. The optical conductivity $\sigma(\omega)$ estimated from the reflectivity data showed two characteristic features that are closely related to the thermoelectric properties of Ta$_4$SiTe$_4$. One is a small energy gap of 0.1–0.15 eV opening at the Fermi energy $E_F$. The other is an anisotropy in the low-energy region, where $\sigma(\omega)$ parallel to the Ta$_4$SiTe$_4$ chains is an order of magnitude higher than that perpendicular to them.

Single crystals of Ta$_4$SiTe$_4$ were synthesized by crystal growth in the vapor phase. A mixture of a 2:1:2 molar ratio of Ta (Rare Metallic, 99.9%), Si (Kojundo Chemical Laboratory, ≥99.9%), and Te (Rare Metallic, 99.999%) powders was sealed in an evacuated quartz tube with 10–20 mg of TeCl$_4$ powder. The tube was heated to and kept at 873 K for 24 h and at 1423 K for 96 h, and then furnace-cooled to room temperature. This process was performed many times to obtain larger single crystals. Single crystals with a maximum width of 100 μm or more and a length of several millimeter were used for the reflectivity measurements described below. A typical example is shown in Fig. 1(b).

Normal incident reflectivity measurements were performed on the as-grown shiny surface at room temperature, using Fourier-type interferometers (0.02–0.06 eV, DA-8, ABB Bomen and 0.05–2.2 eV, FT/IR6600 IRT-5200) and a grating spectrometer (2–4 eV, MSV-5200) [17]. A typical surface used for the measurements is shown in Fig. 1(c). The reflectivity spectra were measured by linearly polarized light oscillating parallel or perpendicular to the Ta$_4$SiTe$_4$ chains, i.e., the $c$-axis. A Ta$_4$SiTe$_4$ single crystal is easy to bend and break apart under stress, reflecting the nature of the one-dimensional van der Waals crystal. This study always used single crystals that were immediately after taken out from an evacuated quartz tube and confirmed to be free of cleavage and twist. An evaporated Au or Ag film on a glass plate was used as a reference mirror. Reflectivity measurements in the visible to VUV region were performed using synchrotron radiation at the BL3B beamline at UVSOR, Institute for Molecular Science. The spectrum was confirmed to be independent of its place within our spatial resolution. For quantitative discussion, the optical conductivity $\sigma(\omega)$ was deduced from the reflectivity $R(\omega)$ by the Kramers–Kronig transformation. This transformation requires appropriate extrapolations. An extrapolation below 0.02 eV was made according to the Hagen–Rubens equation and one above 30 eV assuming $R \propto \omega^{-4}$. Parameters in the Hagen-Rubens extrapolations $\sigma(0)_{HR}$ were 180 and 38 Ω$^{-1}$ cm$^{-1}$ for the $R(\omega)$ spectra taken parallel and perpendicular to the $c$-axis, respectively.

First-principles density functional calculations were performed on Ta$_4$SiTe$_4$ using the Quantum ESPRESSO code [18,19]. The calculations used norm-conserving pseudopotentials from the optimized norm-conserving Vanderbilt pseudopotential library [20] sourced from PseudoDojo [21]. The exchange-correlation function was treated within the generalized gradient approximation in the Perdew–Burke–Ernzerhof formalism [22]. The plane-wave energy cutoff for the wave functions was set to 92 Ry. Brillouin zone integrations were performed on a 4×2×9 k-point mesh. Electronic occupations were smeared with a Gaussian width of 0.002 Ry. In these calculations, spin–orbit coupling was explicitly considered. The results of the density functional calculations were used to calculate the optical conductivity spectra using RESPACK code [23,24]. In RESPACK, optical conductivity is derived from the dielectric function based on a random phase approximation. The energy cutoff of the dielectric function was set to 3 Ry. The integral over the Brillouin zone was calculated using the generalized tetrahedron technique, with a smearing of 0.01 eV.

Figure 2 shows the reflectivity spectra of a Ta$_4$SiTe$_4$ single crystal measured at room temperature using linearly polarized light oscillating parallel to the $c$-axis (an electric field $E$ is parallel to the $c$-axis, i.e., $E//c$) and oscillating perpendicular to the $c$-axis ($E \perp c$). The reflectivity for $E//c$, $R_{//}(\omega)$, shows a peak at around 1.4 eV and a strong increase below 0.7 eV. The latter increase most likely corresponds to the plasma edge in metals. However, $R_{//}(\omega)$ is considerably suppressed and decreased below 0.3 and 0.17 eV, respectively, in the mid-infrared region, followed by a sharp increase below 0.03 eV. This complex spectral shape below the edge at 0.4 eV suggests that Ta$_4$SiTe$_4$ is not a simple metal with Fermi surfaces, but has a complex structure in its band structure near $E_F$, as will be discussed later. The $E \perp c$ reflectivity, $R_\perp(\omega)$, also showed a decrease below 0.17 eV and a sharp increase below 0.03 eV, similarly to $R_{//}(\omega)$. However, the $R_\perp(\omega)$ values are lower than $R_{//}(\omega)$ in the whole energies, and the increase in $R_\perp(\omega)$ below 0.4 eV is much weaker than that in $R_{//}(\omega)$ below 0.7 eV, probably reflecting an anisotropy in electrical conduction.

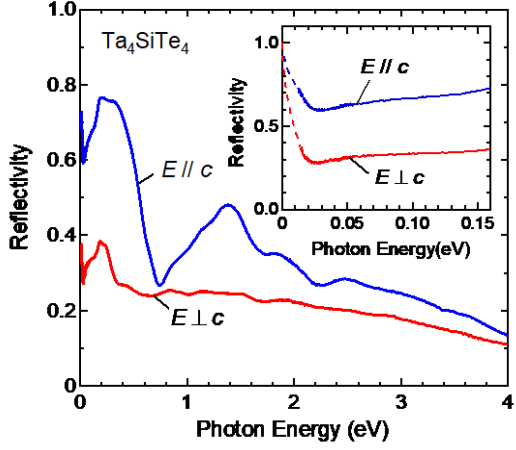

Figure 2. Optical reflectivity spectra of a Ta$_4$SiTe$_4$ single crystal for polarization parallel and perpendicular to the $c$-axis measured at room temperature. The inset shows the enlarged spectra below 0.15 eV. The dotted lines are extrapolations using the Hagen–Rubens formula.

The optical conductivity spectra of Ta$_4$SiTe$_4$ at room temperature obtained by performing Kramers–Kronig transformation of the extrapolated reflectivity spectra are shown in Fig. 3(a). The optical conductivity for $E//c$, $\sigma_{//}(\omega)$, has peaks at 2.5, 1.9, 1.3, and 0.2 eV. The first three peaks in the near-infrared to visible region most likely correspond to the interband transition. The last peak at 0.2 eV corresponds to the band gap at $E_F$, which will be discussed in detail below. In contrast, the optical conductivity for $E \perp c$, $\sigma_\perp(\omega)$, also has a small peak at around 0.2 eV, but does not show a clear peak due to the interband transition. These behaviors agree well with the theoretical result shown in Fig. 3(b). The theoretical spectrum for $E//c$ has four prominent peaks at 2.5, 2.0, 1.4, and 0.4 eV, which correspond to those observed in the experimental spectrum shown in Fig. 3(a). The theoretical spectra of $E//a$ and $E//b$ are almost identical and have no significant structure other than a strong decrease below 0.2–0.4 eV, which is consistent with the experimental $\sigma_\perp(\omega)$. The experimental $\sigma(\omega)$ of Ta$_4$SiTe$_4$ also satisfies the summation rule. The inset of Fig. 3(a) shows the effective electron number per formula unit, $N_{eff}$, which is calculated as $N_{eff} = \frac{2m_0 V}{\pi e^2} \int_0^\omega \sigma(\omega') d\omega'$, where $m_0$ and $V$ are the bare electron mass and the volume per formula unit, respectively. At sufficiently high energies, $N_{eff}$ values for both $E//c$ and $E \perp c$ converge to 48, which is the number of valence electrons in a formula unit of Ta$_4$SiTe$_4$. These results indicate that the spectral analysis performed in this study, including the extrapolations, is appropriate.

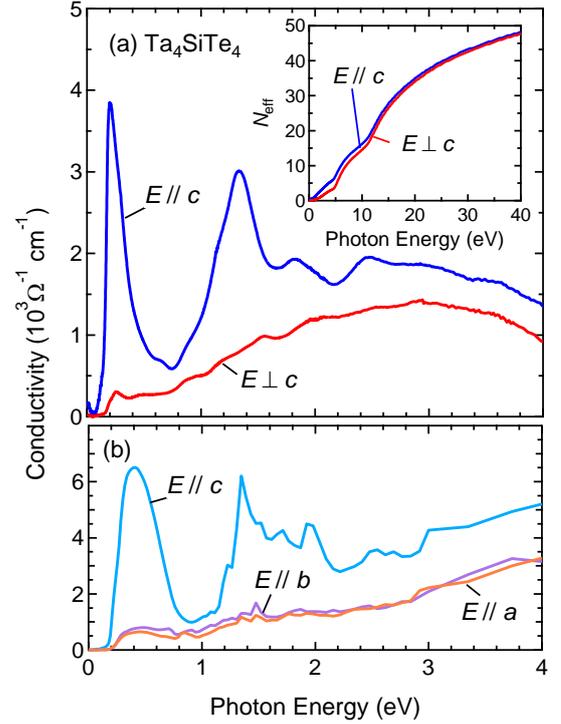

Figure 3. Optical conductivity spectra of a Ta$_4$SiTe$_4$ single crystal parallel and perpendicular to the $c$-axis at room temperature. The inset shows the effective number of electrons per formula unit. (b) Calculated optical conductivity spectra of Ta$_4$SiTe$_4$ parallel to the $a$-, $b$-, and $c$-axes.

Next, the low-energy $\sigma(\omega)$ of Ta$_4$SiTe$_4$, which is closely related to its thermoelectric properties, will be discussed. As shown in Fig. 4(a), $\sigma_{//}(\omega)$ strongly increases from low values above 0.1 eV with increasing $\omega$. The $\sigma_\perp(\omega)$ also increases above 0.15 eV, even though the change is weaker than that for $E//c$. The direct-independent increases of $\sigma(\omega)$ above 0.1–0.15 eV indicate the presence of a small energy gap of 0.1–0.15 eV at $E_F$. The presence of such a small energy gap was implied in the Arrhenius plot of the electrical conductivity of a Ti-doped whisker, where the carrier density was reduced by Ti doping [13]. The small band gap was also pointed out in the first-principles calculations [8,16]. The first-principles calculations without spin–orbit coupling showed that Ta$_4$SiTe$_4$ is a Dirac semimetal with band crossing points at $E_F$. When spin–orbit coupling is switched on, a small band gap of ~0.1 eV opens at $E_F$. The strong increase in $\sigma(\omega)$ above 0.1–0.15 eV is a direct observation of a spin–orbit gap opening at $E_F$ in Ta$_4$SiTe$_4$.

The existence of a small energy gap at $E_F$ is closely related to the high thermoelectric performance at low temperatures of Ta$_4$SiTe$_4$. According to the 10 $k_B T$ rule for thermoelectric materials, there is a relationship $\Delta \sim 10 k_B T_{max}$ between the size of the band gap $\Delta$ and the optimum temperature for a thermoelectric material $T_{max}$ in semiconductor thermoelectric materials [25]. In fact, Bi$_2$Te$_3$, PbTe, and SiGe with band gaps of 0.3, 0.5, and 0.7 eV have $T_{max}$ of approximately 300, 500–700, and over 1000 K [18]. The observed $\Delta \sim 0.1$ eV in Ta$_4$SiTe$_4$ suggests high thermoelectric performance at around 100 K in this material. It is not easy for a material to have such a very small $\Delta$ at $E_F$, which is one of the reasons why thermoelectric conversion has not been put to practical use at low temperatures. For example, CsBi$_4$Te$_6$, which exhibits optimum performance at around 200 K, has a very small $\Delta$ due to the Bi–Bi bonds sparsely existing in its crystal structure [26]. In Ta$_4$SiTe$_4$, a spin–orbit gap opening at the Dirac point gives rise to a small $\Delta$. The realization of a very small band gap in some way can serve as a guideline for finding a promising material for low-temperature applications, which has yet to be put into practical use. This discussion indicates that the spin–orbit gap is a promising approach.

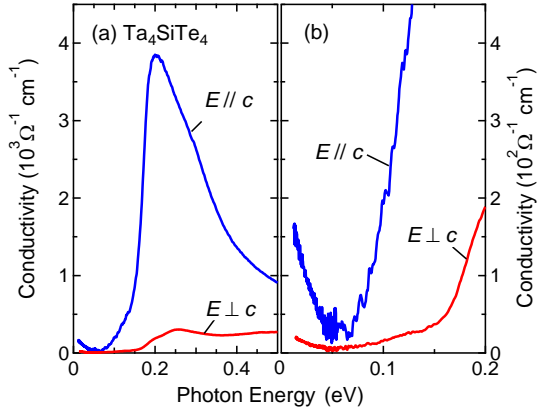

Figure 4. Low-energy optical conductivity spectra of a Ta$_4$SiTe$_4$ single crystal parallel and perpendicular to the $c$-axis at room temperature. (a) and (b) show the optical conductivity spectra below 0.5 and 0.2 eV, respectively.

Below 0.05 eV, $\sigma(\omega)$ of Ta$_4$SiTe$_4$ for both $E//c$ and $E \perp c$ gradually increases toward the lowest energy, as shown in Fig. 4(b), which is probably the Drude peak associated with the itinerant electrons. As a result, $\sigma_{//}(\omega)$ at the lowest energy of 0.02 eV is equal to 160 $\Omega^{-1}$ cm$^{-1}$, which is approximately one-sixth of the dc electrical conductivity of 1000 $\Omega^{-1}$ cm$^{-1}$ measured using a whisker crystal [8]. Although the origin of this discrepancy is not fully understood, $\sigma(\omega)$ may continue to increase below 0.02 eV toward $\hbar\omega = 0$, resulting in a smaller discrepancy between $\sigma(0)$ and the dc conductivity. In this case, the increase in $\sigma(\omega)$ toward $\hbar\omega = 0$ indicates the presence of a small amount of strongly light conducting carriers. This is one of the possible scenarios that can be naturally expected from the fact that Ta$_4$SiTe$_4$ has Dirac-like band dispersions around $E_F$.

Although the quantitative estimate of the lowest-energy $\sigma(\omega)$ remains ambiguous, as discussed above, $\sigma_\perp(\omega) = 20$ $\Omega^{-1}$ cm$^{-1}$ at $\hbar\omega = 0.02$ eV is one-eighth of $\sigma_{//}(\omega)$, as shown in Fig. 4(b), suggesting the presence of an anisotropy of about one order of magnitude in the electrical conduction of Ta$_4$SiTe$_4$. Although this anisotropy seems to be weak, considering that Ta$_4$SiTe$_4$ is a one-dimensional van der Waals crystal, this result clearly demonstrates the presence of anisotropic electron conduction in Ta$_4$SiTe$_4$.

The significance of the observed anisotropy in the electrical conduction of Ta$_4$SiTe$_4$ for its high thermoelectric performance will now be discussed. It has been theoretically shown that materials with anisotropic electrical conduction exhibit higher thermoelectric performance than isotropic materials. An example is the B factor, expressed by (1), giving the upper limit of $ZT$ [25–27]:

$$B = \frac{1}{3\pi^2}\left(\frac{2k_B T}{h^2}\right)^{3/2}\sqrt{m_x m_y m_z}\,\frac{k_B T \mu_z}{e\kappa_{lat}}, \quad (1)$$

where $m_x$, $m_y$, and $m_z$ are the effective masses in the $x$, $y$, and $z$ directions, respectively, $\mu_z$ and $\kappa_{lat}$ are the mobility and the lattice thermal conductivity along the $z$ direction, respectively, and $z$ is the direction in which thermoelectric performance is evaluated. Equation (1) reflects the fact that $S$ depends on the average effective mass for all directions, whereas the electrical conductivity is proportional to $\mu_z$. $B$ is proportional to $\sqrt{(m_x m_y/m_z)}$, assuming that $\mu_z = e\tau_z/m_z$, where $\tau_z$ is the relaxation time of the conducting carriers along the $z$-direction. Therefore, a material with one-dimensional anisotropy with $m_x = m_y > m_z$ has a larger $B$ and can exhibit higher thermoelectric performance than an isotropic material with $m_x = m_y = m_z$ or a two-dimensional material with $m_x > m_y = m_z$. As discussed in the previous paragraph, Ta$_4$SiTe$_4$ has anisotropic electrical conduction indicated by its one order of magnitude higher $\sigma_{//}(\omega)$ than $\sigma_\perp(\omega)$ at the lowest energy. Given that the scattering mechanisms are similar for $//c$ and $\perp c$, as suggested by the fact that both $\sigma_{//}(\omega)$ and $\sigma_\perp(\omega)$ increase below 0.05 eV, the observed anisotropy between $\sigma_{//}(\omega)$ and $\sigma_\perp(\omega)$ is mainly due to the difference in the effective mass, i.e., $m_{//} < m_\perp$. Therefore, the enhancement of thermoelectric performance by the anisotropic effective mass discussed above is expected to be involved in Ta$_4$SiTe$_4$.

In general, however, one-dimensional electron conduction is weak against disorder. External factors such as lattice defects have a significant negative effect on one-dimensional electron conduction due to Anderson localization. If we discuss using the $B$ factor discussed above, $\sqrt{(m_x m_y/m_z)}$ can be enhanced by one-dimensional anisotropy, but $\mu_z$ tends to be strongly suppressed at the same time, meaning that it is difficult to achieve large $B$ in such a material. In contrast, electron conduction in Ta$_4$SiTe$_4$ is robust against disorder, which was clearly shown for small $\rho$ in the Ta$_4$SiTe$_4$–Nb$_4$SiTe$_4$ solid solution samples [12]. It is expected in Ta$_4$SiTe$_4$ that the weak anisotropy of electrical conduction plays an important role in maintaining such a one-dimensional anisotropy. As a result, Ta$_4$SiTe$_4$ maintains high electrical conductivity accompanied by anisotropic electrical conduction with $m_{//} < m_\perp$, which most likely leads to the high thermoelectric performance of this material. Recently, some materials with one-dimensional crystal structure have been found to exhibit high thermoelectric performance below room temperature [28–31], implying the importance of the anisotropic electrical conduction.

In conclusion, we measured the reflectivity of synthesized single crystals of a low-temperature thermoelectric material, Ta$_4$SiTe$_4$, over a wide energy range. The optical conductivity data estimated from the reflectivity spectra indicated the presence of a small band gap of 0.1–0.15 eV at $E_F$, corresponding to the spin–orbit gap predicted in the first-principles calculations. At the lowest measured energy, $\sigma_{//}(\omega)$ is approximately one order of magnitude higher than $\sigma_\perp(\omega)$, indicating the presence of an anisotropy in the electrical conduction of Ta$_4$SiTe$_4$. This very small band gap and the weak but robust anisotropy in electrical conduction play key roles in high thermoelectric performance in Ta$_4$SiTe$_4$ below room temperature. The coexistence of these two factors in a material is expected to be a promising strategy to develop a practical material for low-temperature applications.

**Acknowledgments**

We are grateful to Y. Yoshikawa and Y. Abe for their help with single crystal growth and K. Nakamura for his help with first principles calculations. A part of this work was performed at the BL3B of UVSOR Synchrotron Facility, Institute for Molecular Science (IMS program 21-637). This work was partly supported by JSPS KAKENHI (Grant Nos. 19H05823, 20H00346, 20H02603, 22H04953, and 23H01831) and the Research Foundation for the Electrotechnology of Chubu. A part of the computation was done at Supercomputer Center, Institute for Solid State Physics, the University of Tokyo.

**References**

[1] K. Biswas, J. He, I. D. Blum, C.-I. Wu, T. P. Hogan, D. N. Seidman, V. P. Dravide, and M. G. Kanatzidis, Nature **489**, 414 (2012).

[2] K. F. Hsu, S. Loo, F. Guo, W. Chen, J. S. Dyck, C. Uher, T. Hogan, E. K. Polychroniadis, and M. G. Kanatzidis, Science **303**, 818 (2004).

[3] L.-D. Zhao, S.-H. Lo, Y. Zhang, H. Sun, G. Tan, C. Uher, C. Wolverton, V. P. Dravid, and M. G. Kanatzidis, Nature **508**, 373 (2014).

[4] C. Chang, M. Wu, D. He, Y. Pei, C.-F. Wu, X. Wu, H. Yu, F. Zhu, K. Wang, Y. Chen, L. Huang, J.-F. Li, J. He, and L.-D. Zhao, Science **360**, 778 (2018).

[5] W. M. Yim and A. Amith, Solid State Electron. **15**, 1141 (1972).

[6] D.-Y. Chung, T. Hogan, P. Brazis, M. Rocci-Lane, C. Kannewurf, M. Bastea, C. Uher, and M. G. Kanatzidis, Science **287**, 1024 (2000).

[7] D.-Y. Chung, T. P. Hogan, M. Rocci-Lane, P. Brazis, J. R. Ireland, C. R. Kannewurf, M. Bastea, C. Uher, and M. G. Kanatzidis, J. Am. Chem. Soc. **126**, 6414 (2004).

[8] T. Inohara, Y. Okamoto, Y. Yamakawa, A. Yamakage, and K. Takenaka, Appl. Phys. Lett. **110**, 183901 (2017).

[9] M. E. Badding and F. J. DiSalvo, Inorg. Chem. **29**, 3952 (1990).

[10] J. Li, R. Hoffmann, M. E. Badding, and F. J. DiSalvo, Inorg. Chem. **29**, 3943 (1990).

[11] Y. Okamoto, T. Wada, Y. Yamakawa, T. Inohara, and K. Takenaka, Appl. Phys. Lett. **112**, 173905 (2018).

[12] Y. Yoshikawa, T. Wada, Y. Okamoto, Y. Abe, and K. Takenaka, Appl. Phys. Express **13**, 125505 (2020).

[13] Y. Okamoto, Y. Yoshikawa, T. Wada, and K. Takenaka, Appl. Phys. Lett. **115**, 043901 (2019).

[14] Q. Xu, S. Qu, C. Ming, P. Qiu, Q. Yao, C. Zhu, T.-R. Wei, J. He, X. Shi, and L. Chen, Energy Environ. Sci. **13**, 511 (2020).

[15] Q. Xu, C. Ming, T. Xing, P. Qiu, J. Xiao, X. Shi, and L. Chen, Mater. Today Phys. **19**, 100417 (2021).

[16] S. Liu, H. Yin, D. J. Singh, and P.-F. Liu, Phys. Rev. B **105**, 195419 (2022).

[17] K. Takenaka, M. Tamura, N. Tajima, H. Takagi, J. Nohara, and S. Sugai, Phys. Rev. Lett. **95**, 227801 (2005).

[18] P. Giannozzi, S. Baroni, N. Bonini, M. Calandra, R. Car, C. Cavazzoni, D. Ceresoli, G. L. Chiarotti, M. Cococcioni, I. Dabo, A. D. Corso, S. de Gironcoli, S. Fabris, G. Fratesi, R. Gebauer, U. Gerstmann, C. Gougoussis, A. Kokalj, M. Lazzeri, L. Martin-Samos, N. Marzari, F. Mauri, R. Mazzarello, S. Paolini, A. Pasquarello, L. Paulatto, C. Sbraccia, S. Scandolo, G. Sclauzero, A. P. Seitsonen, A. Smogunov, P. Umeri, and R. M. Wentzcovitch, J. Phys. Condens. Matter **21**, 395502 (2009).

[19] P. Giannozzi, O. Andreussi, T. Brumme, O. Bunau, M. B. Nardelli, M. Calandra, R. Car, C. Cavazzoni, D. Ceresoli, M. Cococcioni, N. Colonna, I. Carnimeo, A. Dal Corso, S. de Gironcoli, P. Delugas, R. A. DiDtasio Jr., A. Ferretti, A. Floris, G. Fratesi, G. Fugallo, R. Gebauer, U. Gerstmann, F. Giustino, T. Gorni, J. Jia, M. Kawamura, H.-Y. Ko, A. Kokalj, E. Küçükbenli, M. Lazzeri, M. Marsili, N. Marzari, F. Mauri, N. L. Nguyen, H.-V. Nguyen, A. Otero-de-la-Roza, L. Paulatto, S Poncé, D. Rocca, R. Sabatini, B. Santra, M. Schlipf, A. P. Seitsonen, A. Smogunov, I. Timrov, T. Thonhauser, P. Umari, N. Vast, X. Wu, and S. Baroni, J. Phys. Condens. Matter **29**, 465901 (2017).

[20] D. R. Hamann, Phys. Rev. B **88** (2013).

[21] M. J. van Setten, M. Giantomassi, E. Bousquet, M. J. Verstraete, D. R. Hamann, X. Gonze, and G. M. Rignanese, Comput. Phys. Commun. **226**, 39 (2018).

[22] J. P. Perdew, K. Burke, and M. Ernzerhof, Phys. Rev. Lett. **77**, 3865 (1996).

[23] K. Nakamura, Y. Yoshimoto, Y. Nomura, T. Tadano, M. Kawamura, T. Kosugi, K. Yoshimi, T. Misawa, and Y. Motoyama, Comput. Phys. Commun. **261** (2021).

[24] K. Nakamura, Y. Yoshimoto, Y. Nohara, and M. Imada, J. Phys. Soc. Jpn. **79**, 123708 (2010).

[25] G. D. Mahan, *Solid State Physics* (Academic Press, New York,

USA, 1997) Vol. 51, pp. 81-157.

[26] P. Larson, S. D. Mahanti, D.-Y. Chung, and M. G. Kanatzidis, Phys. Rev. B **65**, 045205 (2002).

[27] G. D. Mahan, J. Appl. Phys. **65**, 1578 (1989).

[28] A. Nakano, Y. Maruoka, F. Kato, H. Taniguchi and I. Terasaki, J. Phys. Soc. Jpn. **90**, 033702 (2021).

[29] A. Nakano, A. Yamakage, U. Maruoka, H. Taniguchi, Y. Yasui, and I. Terasaki, J. Phys. Energy **3**, 044004 (2021).

[30] Q. Dong, Y. Huang, L. Zhang, J. Bai, J. Cheng, Q. Liu, P. Liu, C. Li, J. Xiang, J. Wang, B. Ruan, Z. Ren, P. Sun, and G. Chen, Appl. Phys. Lett. **122**, 094104 (2023).

[31] Q. Dong, J. Xiang, Z. Wang, Y. Li, R. Lu, T. Zhang, N. Chen, Y. Huang, Y. Wang, W. Zhu, G. Li, H. Zhao, X. Zheng, S. Zhang, Z. Ren, J. Yang, G. Chen, and P. Sun, Sci. Bull. **68**, 920 (2023).

*E-mail: yokamoto@issp.u-tokyo.ac.jp